# TOWARD AN EFFORT ESTIMATION MODEL FOR SOFTWARE PROJECTS INTEGRATING RISK

S. Laqrichi, D. Gourc, F. Marmier
Université de Toulouse, Mines Albi, Centre Génie Industriel
Route de Teillet, Campus Jarlard, 81013 Albi Cedex 09, France

Abstract
According to a study of The Standish Group International, 44% of software projects cost more and last longer than expected. More accurate the effort estimation is; the better the enterprise gets organized and the more the software project respects the commitments on budget, time and quality. Enhancing the accuracy of effort estimation remains an ongoing challenge to software professionals. Several factors can influence the accuracy of effort estimation, namely the immaterial aspect of information system projects, new technologies and the lack of return on experience. However, the most important factor of cost and delay increase is software risks. A software risk is an uncertain event with a negative consequence on the software project. In this article, we propose a methodology to take into account risk exposure analysis in the effort estimation model. In the literature, this issue is little addressed and few approaches are investigated. In this research work, we first present an overview of these approaches and their limits. Then, we propose an effort estimation model that improves the accuracy of estimation by integrating software risks. We finally apply this model to a case study and compare its results to the results of a classic model.

## 1 INTRODUCTION

Effort estimation is an important activity in the bidding process and throughout the entire software development life cycle. Both project managers and client use effort estimation to predict the effort, duration and cost required to develop their software projects in order to establish contracts. Various effort estimation methods may be used, however they do not produce sufficiently accurate results, hence, approximately 44% of software projects, according to The Standish Group International, fail to meet the commitments on quality, time and cost. Therefore, it is important to identify and analyze factors underlying this failure in order to improve the effort estimation accuracy. In fact, several factors have been identified and addressed in literature such as the choice of estimation technique, the estimator's experience, the precision of user's requirements details, the use of new technologies, etc. However, a very crucial factor is not sufficiently addressed which is software risks. Risks in software projects are numerous and may have a deep negative impact on the projects progress and so on the cost, quality and time.

According to Nguyen and al. software risk must be taken into consideration in two main situations. First, during the software development when faced with a risk situation, the manager has to choose a strategy to treat the risk and to keep project on budget and on time. Then, in the response to a bidding, risks have to be correctly identified and assessed and the strategies adequately chosen to obtain a realistic software project effort and cost estimations [1].

For the first situation, various software risk management approaches and tools have been proposed. They allow project managers to manage risks throughout the development software life cycle. These approaches provide methodologies to identify potential software risks, to measure their possible impacts on the project progress and to choose the adequate risk treatment strategies.

However, for the second situation, taking into account software risks in the effort estimation process is little addressed. In fact, the uncertain and evolutionary nature of software risks makes it difficult to take it into account in the effort estimation process.

This study addresses this issue and proposes a process to take into account software risks in effort estimation and an approach to establish effort estimation model integrating software risks.

The present paper is organized as follows: Section 2 presents literature review on effort estimation process, software risk management approaches, and existing effort estimation approaches taking into account software risks. Section 3 presents the proposed process of effort estimation taking into account risk, then, focuses on the establishment process of the effort estimation model integrating software risks. Section 4 presents a case study of the proposed approach taking into account software risks in effort estimation. The results of the case study show that the proposed effort estimation model taking into account software risks is more accurate than the traditional effort estimation model. Section 5 draws some final conclusions and prospects.

## 2 LITERATURE REVIEW

### 2.1 Estimation in software projects

Estimation in software project aims to predict the development effort, cost and time of a software project. The estimation process is based on the approach shown in Figure 1. This process consists of three activities:
-The estimation or the measure of the functional size
-The determination of the development efforts
-The calculation of the duration, cost and resources required at the global level or by sub-assemblies.

Functional size expresses the size of the software which is derived from the quantification of functional requirements specified by users [ISO/IEC14143]. Functional size can be calculated by several methods and techniques of functional measurement such as FPA (Function Point Analysis) and COSMIC FP (COSMIC Function Point), thus it can be expressed in different units such as function points (FP) and source lines of code (SLOC).

The development effort is a function of the functional size [2], it is expressed in man-hours, man-days or man-months. The development effort estimation makes it possible to predict the number of man-months required to perform the software development project. Several estimation models can be used to estimate development effort based on functional size and other project parameters. Otherwise, the multiplication of functional size with a productivity factor can be used to convert functional size to development effort.

The development effort once estimated, enables to determine the duration, staffing, and cost required for software development because development effort is a function of the duration and staffing. Therefore, duration and staffing can be easily determined either by balancing them, or using duration-staffing curves.

Various effort estimation methods and models that can be used in the estimation process are presented above, they can be classified as expert judgment, parametric models, analogy methods, and machine learning methods.

Expert judgment is a widely used technique in various fields, it is based on the expert intuition and experience drawn from previous executed project. Thus, this technique is subjective especially when performed by a single expert. To mitigate this aspect of subjectivity, expert judgment techniques performed by a group of experts such as wideband Delphi and planning poker were developed.

Parametric and algorithmic models are mainly based on equations expressing the effort as a function of discriminant parameters influencing the effort called effort drivers. Parametric models are established using historical data from complete projects, based on this data, effort drivers that influence strongly the effort are determined then effort estimation equations are expressed.

In analogy methods, the effort required by a project (or a task) is estimated using analogy with executed projects ( or tasks).

Machine learning methods such as case based raisonning and neural network techniques have recently been used in conjunction or as alternative to algorithmic models [3]. These methods are trained using historical data of previous projects to produce accurate estimation by automatically adjusting their parameter to better fit the new project to estimate.

### 2.2 Software project risks

A software risk is an event that may or may not take place and that results in negative consequences on a software project [4][5]. It may affect all aspects of the software project, namely the organization, the technology, the personal, etc. [6].

Risk is defined as a measure of the probability and severity of adverse effects [7]. The multifaceted aspect of software risk makes it difficult to measure it.

Several risk identification methods have been proposed to support project manager in risk management. Many researchers proposed checklists that help project manager in risk identification. Based on a survey of several experienced project managers, Boehm developed a list of ten most important risks in software project [8]. Other methods classify risks into classes according to the project element they affect, such as taxonomy-Based Risk Identification. Taxonomy-Based Risk Identification is established by the Software Engineering Institute (SEI), It regroups risk events in three major classes: Product engineering, development environment, and program constraints. Other researchers classify risks into various dimensions used the multifaceted aspect of software risk. In that way, McFarlan identified three dimensions of software risks that are project size, technology experience, and project structure [9]. Barki and al., based on a data base of 120 projects, classify risks into five dimensions: technological newness, application size, expertise, application complexity and organizational environment [10]. Wallace and al. identify 27 software risks that they classify into six dimensions [11].

After the identification of risk events, the risk assessment process consists in the quantification of the importance of these risk events. It measures and quantifies the degree of importance and criticality of software risk. Various techniques can be used for risk assessment. The majority of methods use both risk impact on the project performance and the probability of occurrence to express the importance of software risk.

The most well-known software risk assessment methods are Boehm's method, SRE, SERIM and DoD. These methods are detailed below.

The Boehm's method focuses on the concept of risk exposure to determine the perceived importance of the risk event at the time of assessment. Risk exposure is defined as the product of the loss probability and the loss magnitude (i.e. impact) of each identified risk. In this method, the loss probability and the loss magnitude is assessed using numerical scale or based on categories with associated numerical value for each category, such as for loss probability: improbable (0.0-0.3), probable (0.4-0.6), or frequent (0.7-1.0). Boehm's method can be used in all the phases of software development, but it doesn't handle generic risk implicitly [8].

The software Risk Evaluation (SRE) method uses the same concept of the risk exposure. The probability of occurrence is assessed based on a scale of one to three, while the impact is determined on the basis of the risk effects on the technical performance, cost, schedule and support. SRE method was developed by SEI, therefore it can be applied to any software project.

Karolak proposed Software Engineering Risk Management (SERIM) method based upon Just-In-Time (JIT) strategy. The JIT software aims to minimize risks and their contingencies, and to manage the risks early in software life cycle. The SERIM associates to each software risk a specific metric and a question. These questions form a checklist that enables the user to identify software risks.

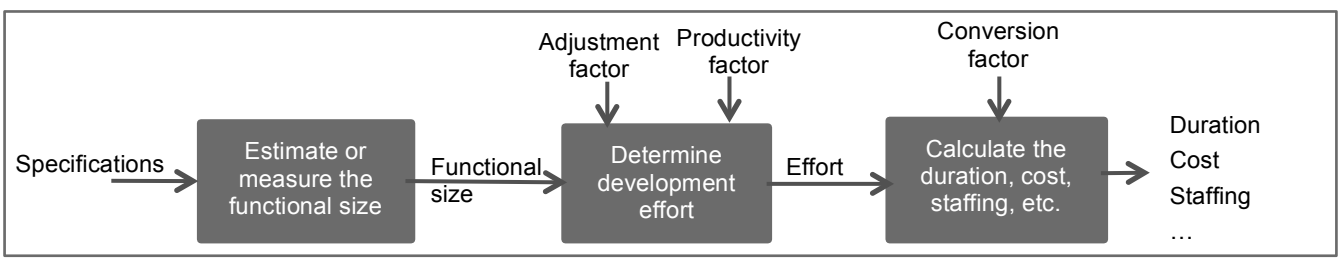

Figure 1: Generic software estimation process.

The answers to the questions are then converted to numerical values through metrics and used to calculate risk factor values, using probability tree [12].

The Department of Defense and the Defense Acquisition University (DAU), proposed the Risk management Guide for the Department of Defense (DoD) to assist software project managers in identification, assessment and management of risks during the entire software life cycle. In this method, probability is assessed using a scale of 5 levels, namely "remote", "Unlikely", "Likely", "Highly likely" and "Near certainty", whereas impact is identified in terms of technical performance, schedule, cost and team [13].

### 2.3 Risk in effort estimation

According to a study of The Standish Group International, 44% of software projects cost more and last longer than expected. This may be the result of inaccurate and unrealistic effort estimation.

Since the risk is a major factor that prevents software project from being on time and staying within budget, it is important to take it into account during the estimation process.

Several approaches were proposed to handle this issue such as the simplistic approach, the Function Point Analysis, COCOMO II, etc.

In many cases, project managers use the simplistic approach that consists on applying a risks factor to the estimated effort, duration, and cost in order to integrate risk in estimation. The risks factor is independent of the used estimation method; the project manager, based on his experience, determines it. It ranges from 1 to *x*. That means that if the factor is 1, no risk will occur, and when the factor is *x*, all risks will occur. This approach is not complicated to handle, but it doesn't allow determining the effect of a specific risk because the risks factor is affected to all project risks without specifying individual risks and assessing each one.

In Function Point Analysis (FPA), it was assumed that project risks are reflected in the fourteen "General System Characteristics (CSC)" and that the overall contingency is expressed in the value adjustment factor. The contingency is defined as the reserve that is set aside to manage and handle the impact of risk events in order to protect projects from producing undesirable results [4]. Like the previous introduced approach, this one doesn't makes it possible to determine the effect of a specific risk due to the unknown relationship between a specific project risk and the general system characteristics.

COCOMO II remedies the underlined lakes of the two previously presented approaches; it defines a risk factor that, based on six types of risk, characterizes each module to be developed [2]. Each risk should be allocated to one risk type whereas a risk can be multidimensional so allocated to different types, which would make it difficult to adjust its treat.

Other approaches were developed to take into account project risks in cost estimation such as contingency estimation model proposed by Uzzafer. In this model, the contingency is defined as the buffer between the expected cost of the project and the expected cost due to the maximum impacts of risk events. Therefore, the contingency model provides estimates of man-months reserves to abate the maximum impact of risk. This model is generic and independent of the cost estimation and risk assessment models, and take into account software risks, but it does not provide a unique value of cost estimation and does not take into account the updated software risks predictions.

## 3 APPROACH

In this paper, an approach that integrates software risks in effort estimation process is proposed. The objective of this approach is to assist project manager and estimators in the entire effort estimation process taking into account software risks by providing a methodology and its associated tools and supports.

The effort estimation process integrating risks, as shown in the Figure 2, consists of four steps: the risk identification, the risk analysis, the risk assessment, and effort estimation. Some existing methods, tools and supports are proposed for each step. Methods and tools that are bold in the figure 2 are selected for this study. A focus is given to the effort estimation step for which we propose a new effort estimation model that integrates software risks.

### 3.1 Risk identification and analysis

In the stage of identification, it is important to identify as much as possible software risks that may occur in the software project. That will ensure realistic and accurate software risk measurement. As mentioned in section 2.2, there are several software risks identification methods. For the present approach, the Wallace and al. software risk structure is used (Table 1), because it is a rigorous validated identification's method that is grounded in the IS literature and validated with practicing software project managers. Besides, it provides an exhaustive risks' list that are classified in six dimensions. Furthermore, in this classification scheme the categories or dimensions of risk are as distinct as possible.

### 3.2 Risk assessment

Our approach aims through this step to assess the software risks for the entire project and to determine the project total risk exposure.

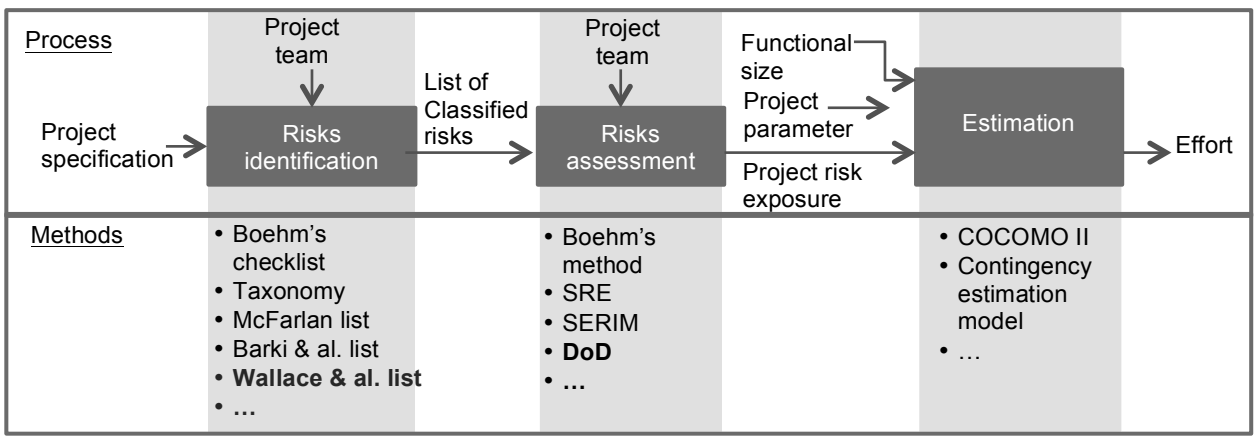

Figure 2: Process of effort estimation integrating risks and examples of support methods.

For this purpose, the DoD risk assessment method is selected based on the comparison of the various impacts for the software assessment methods mentioned in section 2.2. The comparison made by Han and Huang shows that compared to other methods, the DoD method contains diverse types of impacts and describes clearly each assessment criteria and each probability scale of occurrence and impacts [15].

Table 1: Software risks list [11].

| Risk dimension | Software risk |
| --- | --- |
| User | Users resist to change |
| | Conflict between users |
| | Users with negative attitudes toward the project |
| | Users not committed to the project |
| | Lack of cooperation from users |
| Requirement | Continually changing system requirements |
| | System requirements not adequately identified |
| | Unclear system requirements |
| | Incorrect system requirements |
| Project complexity | Project involved the use of new technology |
| | High level of technical complexity |
| | Immature technology |
| | Project involves use of technology that has not been used in prior projects |
| Planning & control | Lack of effective project management methodology |
| | Project progress not monitored closely enough |
| | Inadequate estimation of required resources |
| | Poor project planning |
| | Project milestones not clearly defined |
| | Inexperienced project manager |
| | Ineffective communication |
| Team | Inexperienced team members |
| | Inadequately trained development team members |
| | Team members lack specialized skills required by the project |
| Organizational environment | Change in organizational management during the project |
| | Corporate politics with negative effect on project |
| | Unstable organizational environment |
| | Organization undergoing restructuring during the project |

In this paper, the project total risk exposure is defined as the average of dimensions risks exposure. For each dimension in the Table 1, the manager calculates the risk exposure for each software risk, and then calculates the average of these risks exposure.

The risk exposure is computed by multiplying the probability of occurrence with the composite impact.

The probability of occurrence is assessed based on the manager's and team experience and using the Table 2, it can take a value from 1 to 5.

The composite impact of a software risk is the average of the four components that are: impact on the technical performance, impact on the cost, impact on the schedule and impact on the team. These components are assessed based on the managers and teams experience and using the Table 2; a value from 1 to 5 is assigned to each component depending on its level.

### 3.3 Effort estimation integrating risk

This paper proposes a framework, shown in Figure 3, for establishing effort estimation models taking into account project software risks and based on database of completed projects. This framework consists of four stages. First, the parameter selection stage aims to define the discriminant effort drivers of the project development effort based on a data set. These effort drivers will be used to build effort estimation models. Next, in the data preparation stage, completed projects database is prepared using effort drivers outcomes from the parameter selection stage in addition to assessed projects risks. Afterwards, the effort estimation models integrating risks establishment can take place. Several accuracy indicators are finally used to validate the effort estimation model.

*Parameter selection*

In this stage, a projects database is explored to extract the most discriminant effort drivers. Organization's projects database is built over years by projects teams in order to capitalize the experience and information related to completed projects. It contains information about complete projects such as development effort, functional size, and platform development. In order to establish an effort estimation model taking into account software project risk, project risk exposure is assessed and updated by project managers through return of experience at the end of each project.

To identify the most relevant effort drivers, the statistical test of Pearson correlation and one-way ANOVA can be used [16], they enables to examine the significance between the effort drivers and the development effort in order to select the effort drivers with significant influence on the development effort.

The Pearson's correlation test is used for the effort drivers with the ratio scale to investigate the correlation of two datasets [17]. It provide a correlation coefficient ranging from -1 to 1, it takes value 1 if the two datasets are perfectly correlated, 0 if they are completely uncorrelated, and -1 if they are perfectly anti-correlated.

One-Way Analysis of Variance (ANOVA) test allows determining if one given effort driver has a significant effect on the distribution of the development effort. It is used for effort drivers with the nominal scale.

The relevance and the influence of project risk exposure (PRE) on the effort are proved using Person correlation test. The application of Pearson correlation test on the project risk exposure of a sample of 164 projects provide a correlation coefficient of 0.43, hence it shows that risk exposure is considered as a relevant effort driver.

*Data preparation*

In this stage, the projects database is prepared in order to establish effort estimation model. Datasets related to irrelevant effort drivers are deleted. Projects with missing values in effort driver fields are discarded from use in estimation model establishment.

The development effort distribution is adjusted to be normal by discarding atypical projects and projects with asymmetric efforts.

The projects data should be divided into two segments, one used to establish and train the effort estimation model integrating risk and the other used to validate it. The k fold cross-validation approach can be used for this purpose.

Table 2: Risk assessment levels used in the DoD method [13].

| Level | Probability of occurrence | Impact | | | |
|---|---|---|---|---|---|
| | | Technical performance | Cost | Schedule | Team |
| 1 | Not likely (~10) | Minimal or no impact | Minimal or no impact | Minimal or no impact | None |
| 2 | Unlikely (~30) | Acceptable with some reduction in margin | <5% | Additional resources required. Able to meet need dates | Some impact |
| 3 | Likely (~50) | Acceptable with significant reduction in margin | 5-7% | Minor slip in key milestone. Not able to meet need dates | Moderate impact |
| 4 | Highly likely (~70) | Acceptable; no remaining margin | 7-10% | Major slip in key milestone or critical path impacted | Major impact |
| 5 | Near certainty (~90) | Unacceptable | >10% | Cannot achieve key team or major program milestone | Unacceptable |

In this approach the data is divided into k equally or near equally folds or segments. Subsequently k iterations of training and validation are performed such as for each iteration a different fold of the data is used for model validation, while the remaining k-1 folds are used for model training. The value of k is often set to three in literature[18].

*Effort estimation model establishment*

After preparing projects database that specifies for each project the development effort and values of relevant effort drivers and risk exposure, the effort estimation model can be established using regression.

There are various types of regression that have been used in effort estimation models, namely linear or multi linear regression like the one used by kok and al. [19], non-linear regression that was used by Boehm [20], and ordinal regression that was used by sentas and al. [21].

Project database may contain quantitative effort drivers with ratio scale and also qualitative effort drivers with nominal scale. Thus, effort estimation model can't be directly derived from database by applying usual multi linear regression (MLR) that is sensitive to the categorization of variables. In order to address this issue, different manner can be used such as the Analysis of Covariance (ANCOVA) and the Ordinal Regression (OR).

ANCOVA is a regression model that contains quantitative and qualitative variables that are called dummies. Dummies are categorical variables that take the value of either 0 or 1 indicating the absence or the presence of a categorical effect on the outcome [22].

The Ordinal Regression is a generalization of the MLR predicting cumulative probabilities for the ordered categories of the dependent variable. It provides a separate equation for each category and each equation provides us with a predicted probability of being in the corresponding category or any lower category [21].

Besides these traditional types of regression, the recent machine learning approach can be used to establish a learning based effort estimation model integrating risk. Learning based models are capable of learning Incrementally as new data are provided over time [23], thus, they are automatically adjusted and adapted to the new project to estimate. Well known ML approaches are artificial neural networks (ANN), Support vector regression (SVR), Regression Trees, and Multiple additive Regression Tree (MART) etc.

*Model validation*

In order to compare performances and accuracies of traditional effort estimation model not taking into account risks and effort estimation model integrating risk, different accuracy indicators can be used for this study such as the Mean Magnitude of Relative Error (MMRE), the Pred(0.25) and the coefficient of determination $R^2$.

Both the Mean Magnitude of Relative Error (MMRE) and the Pred(0.25) statistics were suggested by conte and al. [24]. The MMRE is based on the calculation of the Magnitude of Relative Error (MRE) that is a percentage of the estimation error in comparison with the actual development effort used by the executed project as shown in Equation (1) where $y_j$ is the actual effort of the project $j$ and $\hat{y}_j$ is the effort estimation of the project $j$ [25].

$$MRE_j = \frac{|y_j - \hat{y}_j|}{y_j} \quad (1)$$

The MMRE is then the mean of the MREs of set of software project [26].

Pred is a measure of the predictive ability of an effort estimation model where Pred(0.25) compute the percentage of projects whose MRE is less than or equal to 25% [26].

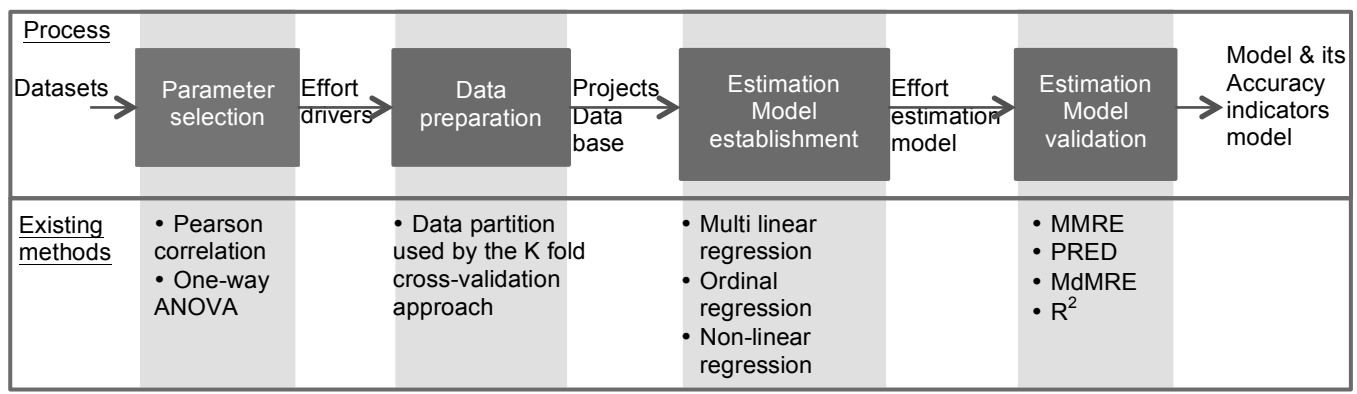

Figure 3: Generation process of effort estimation models.

The coefficient of determination measures the correlation between effort estimates and actual effort based on the idea that bigger projects generate bigger estimates than smaller projects. Albrecht [27] and others have used linear regression to measure this correlation. The coefficient of determination ranges from 0 for no correlation to 1 in the case of the perfect correlation. For an effort estimation model, the coefficient of determination is measured; the higher it, the better the model [28].

## 4 CASE STUDY

The database used in this case study contains 234 projects with 55 data items for each project structured in the same way as ISBSG (International Software Benchmarking Standards Group) in addition to the project risk exposure. The project risk exposure is assessed using the method detailed above.

### 4.1 Parameter selection

Based on the analysis performed by Huang and al. on the ISBSG repository containing 1238 projects, eight effort drivers are considered in this study: Functional size (FS), max team size (MTS), development type (DT), development platform (DP), language type (LT), used methodology (UM), methodology acquired (MA), and application type (AT). FS, MTS, and PRE are quantitative effort drivers, while DT, DP, LT, UM, MA, and AT are qualitative effort drivers.

The Person correlation test and the one-way ANOVA described in 3.3 are adopted to test the relevance of these effort drivers based on the database of this study. As shown in Table 3, in addition to PRE four of these effort drivers display important influence that are: FS and MS with ratio scale, and LT and MA with nominal scale. These effort drivers are thus included in the effort estimation model establishment in addition to PRE.

Table3: Results of Pearson correlation and one-way ANOVA.

| Correlation Test | Effort driver | Correlation |
|---|---|---|
| Pearson | FS | 0.247 |
| | MTS | 0.264 |
| | PRE | 0.436 |
| One Way ANOVA | DT | 0.002 |
| | DP | 0.002 |
| | LT | 0.03 |
| | MA | 0.01 |
| | UM | 0.006 |
| | AT | 0.001 |

### 4.2 Data preparation

After discarding datasets related to irrelevant effort drivers then projects with missing values and atypical projects with outliers, the database is reduced to 168 projects.

The three fold cross-validation is used in this study, thus the database is divided into three equally folds, two of them are used to establish effort estimation model integrating risk while the remaining fold is used to validate the model.

### 4.3 Effort estimation model integrating risks establishment

In this study, effort estimation model integrating risks is derived from the two-thirds of the database. Effort is expressed in terms of the four effort drivers in addition to the project risk exposure using the multi linear regression.

In the same time, a traditional effort estimation model is derived from the same database in order to compare results and accuracies in the model validation step. This model is expressed in terms of the four effort drivers using multi linear regression.

### 4.4 Model validation

Afterwards the effort estimation model integrating risk establishment based on the two-third of the database, the remaining database is used for testing the model and comparing it with the traditional effort estimation model.

Three accuracy indicators are used for this case study: MMRE, Pred(0.25) and $R^2$. Table 4 shows accuracy indicators results in the two phases of training and test of the two effort estimation models: traditional effort estimation model (TEEM) and established effort estimation model integrating risks. In comparison to the traditional effort estimation, the established effort estimation model integrating risks shows better accuracy indicators results for this case study. Therefore, considering the project risks exposure a relevant effort driver and integrating it in the effort estimation model provide significant improvement compared to traditional effort estimation model.

## 5 CONCLUSION

The more accurate effort estimation is, the better the software project complies with the contractual commitments in terms of budget and duration. Crucial factor affecting effort estimation accuracy is software risks, thus, it has to be taken into account in the effort estimation process. This issue is not addressed enough in the literature, that why the present paper handle it.

These research works provide project managers a process to take into account software risks in effort estimation activity. Also, it provides them tools and methods that can be used in each process step.

The process of integrating software risks in effort estimation activity starts by the anticipation and identification of software risks that may occur during the software development, their assessment, then the determination of global project risk exposure. This project risk exposure is considered a relevant effort driver as its influence on the development effort is proved using Pearson correlation test.

Once the project risk exposure is determined, development effort can be estimated using an effort estimation model integrating risks, which is established, based on historical data from previous project. In addition to project risk exposure, important effort drivers are determined, and then the effort estimation model integrating risk is derived from the database.

Table 4: Accuracy comparison of the effort estimation models.

| Model | MMRE | | Pred(0.25) | | $R^2$ | |
|---|---|---|---|---|---|---|
| | Training | Test | Training | Test | Training | Test |
| TEEM | 0.53 | 0.55 | 0.34 | 0.33 | 0.215 | 0.23 |
| EEMR | 0.38 | 0.35 | 0.57 | 0.46 | 0.322 | 0.28 |

The effort estimation model integrating risks, once determined, can be used for new software projects to estimate. However, several model parameters may change over time due to many factors such as the team experience evolution, the team adaptation to technologies, etc. Thus, effort estimation model integrating risks have to be regenerated when the project managers observe that their current model becomes less accurate and the database gets sufficiently bigger.

For the same purpose of improving effort accuracy, future research work can focus on establishing a machine learning based effort estimation model integrating risks. The advantage is that the learning based model is automatically adjusted over time.

During the development software life cycle, effort estimation is refined as requirement becomes more detailed and more accurate. The evolutionary and temporal nature of risks has to be taken into account as well.

Future research studies can focus on the need of continuous software risk assessment and the importance of taking into account the temporal aspect of software risks to refine and update effort estimation through the project.

# 6 ACKNOWLEDGMENT

The research studies presented in this paper are conducted under the project entitled ProjEstimate and financed by the French Governmental Funding (FUI).